\documentclass[usenatbib]{mn2e}
\usepackage{graphicx}
\usepackage{ifthen}
\usepackage{url}

\newcommand{\ltaraw}{$\; \buildrel < \over \sim \;$}
\newcommand{\lta}{\lower.5ex\hbox{\ltaraw}}
\newcommand{\gtaraw}{$\; \buildrel > \over \sim \;$}
\newcommand{\gta}{\lower.5ex\hbox{\gtaraw}}

\newcommand{\sfr}{{\rm\,M_\odot\,yr^{-1}}}
\newcommand{\lsun}{{\rm\,L_\odot}}

\newcommand{\ffffff}[1]{\mbox{$#1$}}
\newcommand{\scnd}{\mbox{\ffffff{''}\hskip-0.3em.}}

\newcommand{\mum}{$\,\mu$m}
\newcommand{\mm}{MM8}

\loadboldmathitalic 
\title [Submm detection of Westphal-MM8]
{Submillimetre detection of the $z=2.83$ Lyman-break galaxy, Westphal\,MM8, and implications for SCUBA2} 
\author[S.\ Chapman \& C.\ M.\ Casey]
{
S.\ C.\ Chapman$^{1,2,3}$, C.\ M.\ Casey$^{1}$\\
$^{1}$ Institute of Astronomy, Madingley Road, Cambridge, CB3 0HA, U.K.\\
$^{2}$ Department of Physics and Astronomy, University of Victoria, Victoria, B.C., V8P 1A1, Canada\\
$^{3}$ Canadian Space Agency, Space Science Fellow\\
}
\date{Accepted 27 February 2009. In original form 26 June 2008.}
\pagerange{\pageref{firstpage}--\pageref{lastpage}} \pubyear{2009}

\begin{document} 
\maketitle
\label{firstpage}

\begin{abstract}
We present confusion limited submillimetre (submm) observations with the SCUBA camera on the JCMT
 of the $z=2.83$ Lyman-break galaxy (LBG), Westphal-MM8, reaching an 850$\mu$m sensitivity even greater than that achieved in the SCUBA map of the Hubble Deep Field region. 
The detection of \mm\ (S$_{850 \mu m} = 1.98\pm0.48$\,mJy), along with the literature submm detections of lensed LBGs, suggest that the LBG population may contribute
significantly to the source counts of submm selected galaxies in 
the 1-2\,mJy regime. Additionally, submm-luminous LBGs are a viable progenitor population for the recently discovered {\it evolved} galaxies at $z\sim2-2.5$.
These observations represent an important baseline for SCUBA2 observations which will regularly map large regions of the sky to this depth.
\end{abstract}




%

\section{Introduction}\label{introduction}

Lyman-break galaxies (LBGs) 
represent the largest population of high-redshift ($z>3$)
star forming galaxies yet discovered (Steidel et al.~1999),
likely comprising a significant percentage of the far-infrared (FIR) background
(Adelberger \& Steidel 2000).
The dust-corrected star formation rate (SFR) distribution of LBGs, n(SFR), shows significant numbers
of $z\sim3$ galaxies with SFR\gta100~M$_\odot$/yr (representing S$_{850 \mu m}\sim1$\,mJy depending on adopted dust parameters).
However, the search for submm counterparts to 
LBGs has proved difficult due to uncertainties in the
relations used to predict the rest frame far-IR luminosity from the UV, 
and also 
photometric errors. 
Since dusty, submm luminous LBGs would characteristically have their UV continuum highly
suppressed due to dust extinction, 
it is difficult to distinguish between 
a truly prodigious star former and a less luminous galaxy with an intrinsically low SFR and 
large photometric errors. 
The FIR and submm
emission from LBGs, their dust content, and their contribution to the
FIR background light remains an unsolved puzzle for which major
progress can only be expected with the new sensitive instruments and
telescopes such as SCUBA-2 and ALMA.

A few highly lensed LBGs have been detected in the submm wavelengths, and represent good test cases for the far-IR properties of $z=3$ star forming galaxies (e.g., Smail et al.\ 2007). However, these objects
could suffer from differential lensing of the UV and far-IR components (e.g., Borys et al.\ 2004)
and may be more difficult to gauge relative to the general population.
The sole submm detection of an unlensed LBG thus far has been for the extremely red ($R-K\sim5$) specimen, Westphal-MMD11 (Chapman et al.~2000). Baker et al.\ (2004) demonstrated through sensitive CO(4-3) observations that it is not a typical member of the submm galaxy population
(c.f., Greve et al.\ 2005). 
While one possibility is that the dust in LBGs is significantly hotter than typical submm galaxies (making them harder to detect at 850$\mu$m), similar to the radio-selected $z\sim2$ ULIRGs in Chapman et al.\ (2004a) and Casey et al.\ (2009), it would be surprising if this characterized the bulk of the LBG population since nothing about their properties suggests they would have hotter dust on average than other classes of star forming galaxies.
Several LBGs have been detected at 24$\mu$m using {\it Spitzer} (Huang et al.\ 2005, Reddy et al.\ 2006), although it is not yet clear whether this excess 24$\mu$m emission comes from a truly high SFR, intricacies in the aromatic dust features, or a power law AGN component. 
Deep submm observations comparable to those presented in this work will be the ultimate diagnostic for star formation in high-$z$ LBGs.

\begin{figure*}
\label{orient}
\begin{center}
\includegraphics[angle=0,width=0.450\hsize]{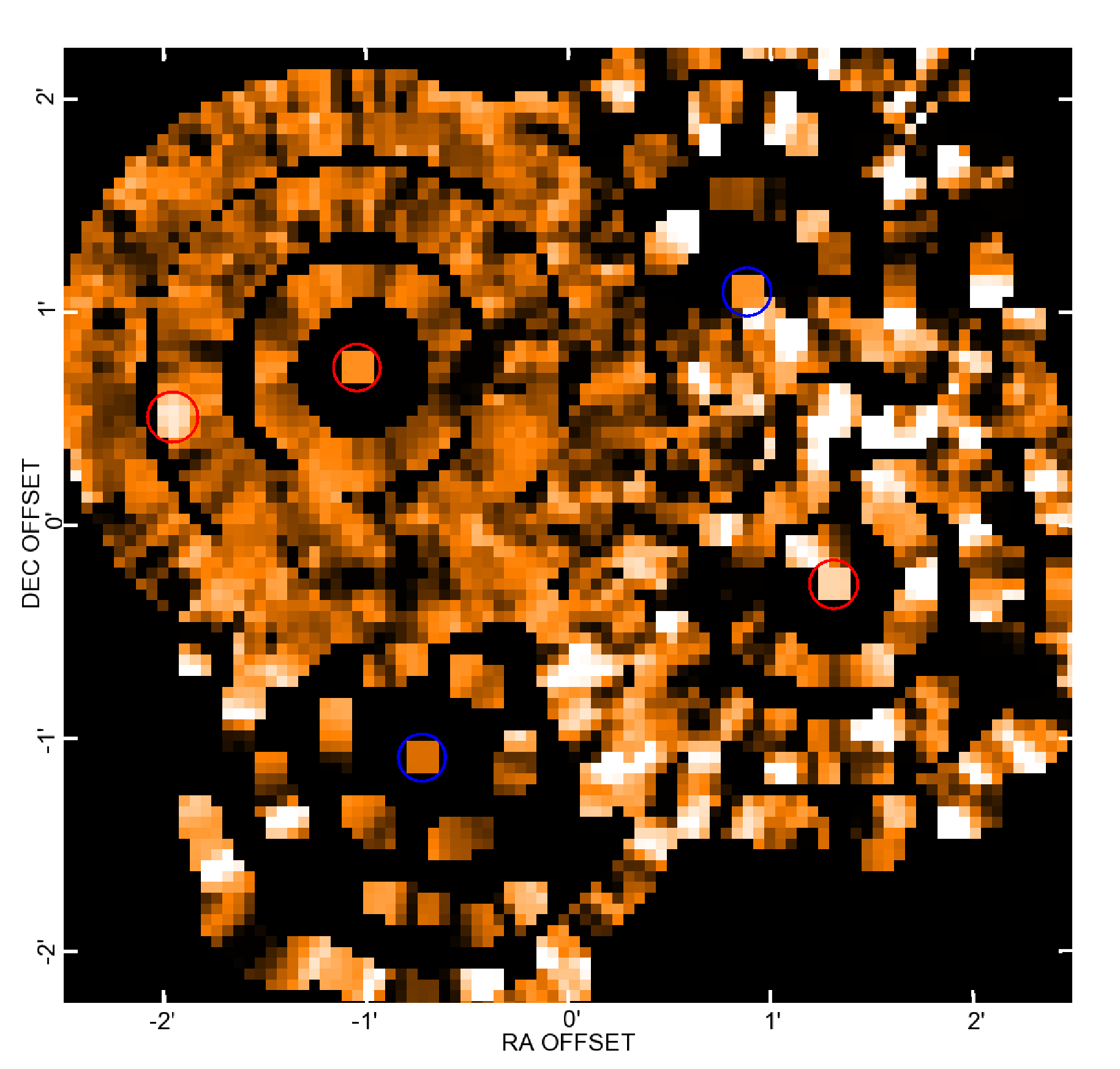}
\includegraphics[angle=0,width=0.450\hsize]{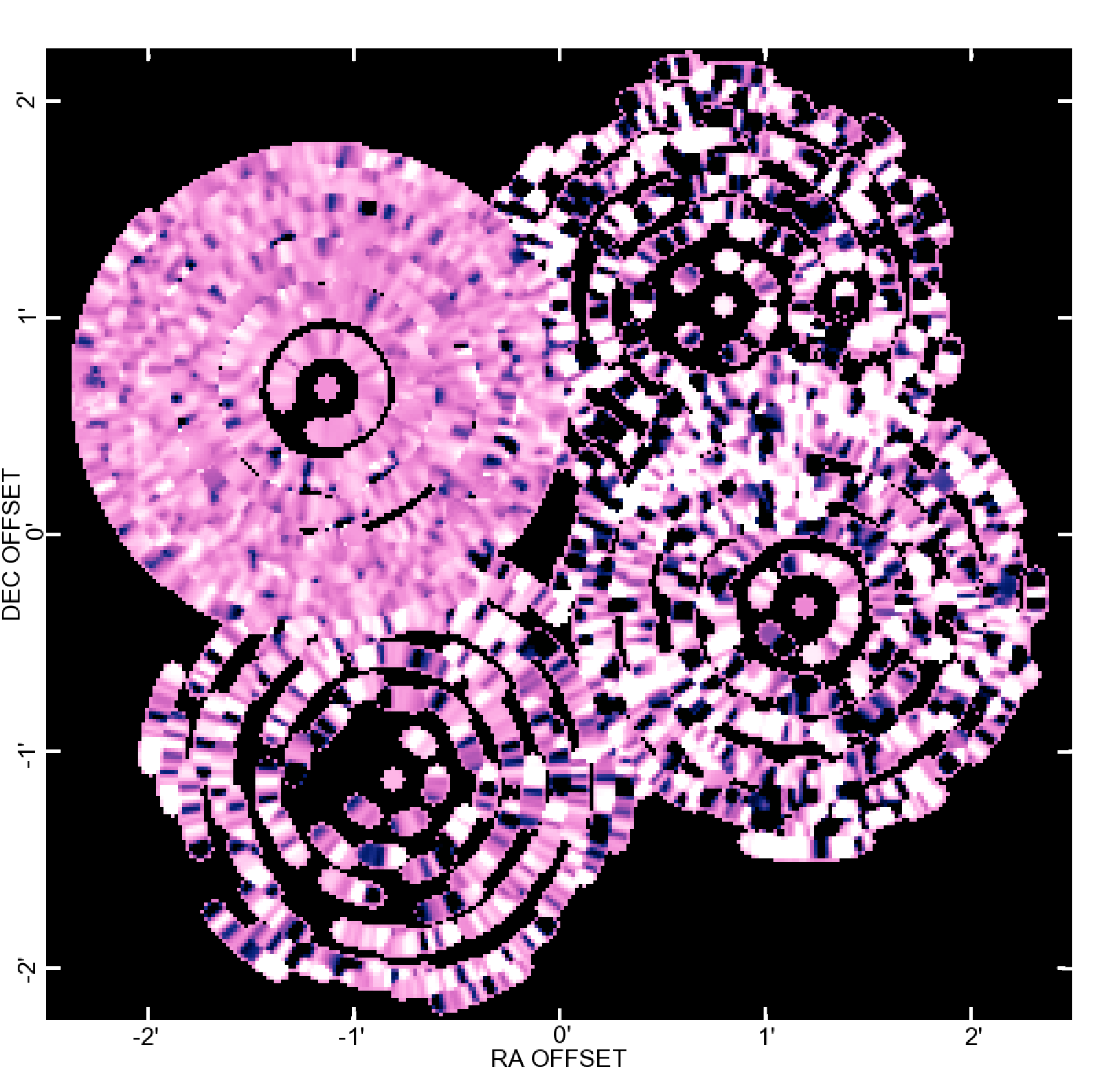}
\caption{SCUBA 850$\mu$m and 450$\mu$m images of the region surrounding the LBGs MM8, MMD11.
The field is 4.5\arcmin$\times$4.5\arcmin, where the 450$\mu$m array covers a slightly smaller area than the 850$\mu$m, but with 191 (versus 37) bolometers.
The two detected LBGs, MM8 and MMD11 are circled in red, as is a serendipitously detected SMG in the deep map surrounding MM8.
}
\end{center}
\end{figure*}

Identified  as a LBG in the surveys of Steidel et al.~(2003), Westphal-MM8 
has from its  rest-frame  UV  spectrum 
an exceptionally large SFR ($>100 \sfr$, after correction for dust extinction).
However, the infrared colors
and overall properties do not single this object out as unusual
for the population in any other respects.
In  this letter we  present deep confusion limited submm  observations  of \mm,
probing  the  emission from the dusty component.
Throughout we assume a cosmology with $h=0.7, \Omega_\Lambda = 0.72, \Omega_M = 0.28$ (e.g., Hinshaw et al.\ 2008).

\section{Observations and Analysis}\label{observations}

Westphal-\mm\ was observed  with the  Submillimetre  Common-User Bolometer
Array (SCUBA -- Holland et al.~1999) 
on the James Clerk Maxwell Telescope (JCMT)
on Mauna Kea, Hawaii.
The observations presented here are the combination of the original
observation presented in Chapman et al.~(2000) (5.4\,ks on source) 
and two archival observation (6.7\,ks and 6.0\,ks on source respectively).
The combined observation represents two complete 8\,hr JCMT {shifts}  centered on  this single LBG, and reaches a sensitivity level even deeper than the Hughes et al.\ (1998) SCUBA map of the Hubble Deep Field.

All observations were taken in the standard PHOTOMETRY mode of SCUBA,
divided into $\sim$30\,min scans
with azimuthal chop angle to minimize atmospheric gradients, a chop
frequency of 7.8125\,Hz, and a throw of either 40\arcsec\ or 45\arcsec.
All observations were taken between 1998 and 2000 under superb submm observing conditions.
The sky rotation over the fixed Nasmyth array coordinate 
resulted in all angles on the sky being covered
by the chop throw (Fig.~1). We make no attempt to fold in the few times when the
source actually crossed another bolometer in an offbeam.

Pointing was checked
hourly on blazars and sky-dips were 
performed routinely to measure the atmospheric
opacity directly. The RMS pointing errors were below 2 arcsec, while  the
average atmospheric zenith  opacities at 450\mum\ and 850\mum\ were
1.7 and 0.20 respectively.
The data were reduced using the  Starlink package {\sc SURF} (Scuba User
Reduction Facility, Jenness et al.~1998).
Spikes were first carefully rejected from the
data, followed by correction for atmospheric opacity
and sky subtraction using the median of all the array pixels, except for
obviously bad pixels and the source pixels.
The data were then calibrated against standard planetary and
compact \hbox{H\,{\sc ii}~} region sources, observed  during the
same  nights.

All the scans on source were combined with inverse variance weighting. 
The final signal measured is 1.98$\pm$0.48\,mJy at 850\mum, with no significant detection at 450\mum. 
Dividing the dataset in half shows a similar detection significance in each half,
increasing our confidence that a true detection has been achieved (that the final result is not the
bias of any small part of the dataset).
The \mm\ observation is adjacent to three other shorter exposures of LBGs, included the detected MMD11 described above and two others showing insignificant but positive flux density. We combine all these observations into a single map for illustration and comparison (Fig.~1). 

At such faint SCUBA flux levels, we must be cautious of both 
confusion noise and interloping source effects on our signal.
The probability that a random galaxy lies within the beam is given by
$P = 1 - \exp(-\pi n \theta^2)$, where $n(>S)$ is the
cumulative surface density for the population in question, and
$\theta$ is the beam radius.  For a flux density of $1.9\,$mJy the source
counts are about 5000 per square degree (e.g.~Blain et al.~1999), and so
$P\,{\simeq}\,4$\% per pointing.
Although nine high-SFR LBG candidates were observed in the original
Chapman et al.~(2000) sample, the average sensitivity reached was
only 1.2\,mJy RMS. Thus although the 
chance that at least one of these
observations is contaminated by a $1.9\,$mJy source is then ${\sim}\,27$ 
per cent,
the \mm\ observation is the only one for which any significant
flux would have been measured. In addition, as we shall demonstrate
below, the potential {\it interloper} is highly likely to be drawn from 
the high-SFR LBG population in the first place, making it even more unlikely
that some optically faint source is in fact contributing
this submm flux.

The issue of confusion
noise (Scheuer 1957, Condon 1974) considers 
the fluctuations due to undetected sources which
contribute to our error bars. A full discussion of the effects of
confusion on deep SCUBA photometry observations can be found in
Chapman et al.~(2000).
Blain et al.~(1998) quote a value of $0.44\,$mJy for the variance due
to confusion, derived from their source counts.
This is comparable to our measured RMS for \mm, and since the errors add
in quadrature, this is the dominant source of noise.
The actual flux of \mm\ may therefore be skewed 
slightly brighter than our measured value.
Again, however, we note that the contribution to the confusion noise
from the 1-2\,mJy submm sources is likely to arise from the LBG population
itself, and thus by pointing at a known high-SFR LBG, we have somewhat
reduced the likelihood of confusion bias.
For the other literature, unlensed LBG sources considered in this letter, confusion noise 
is sub-dominant and does not significantly affect our result.

We also searched for additional submm sources significantly detected in the deep maps.
We detect one additional SMG in the outer region of the 850um map (S$_{850 \mu m}$=5.8$\pm$1.4\ mJy), which is marginally detected at 450$\mu$m.
While the outer regions of the map are not as deep as the central point, at $\sigma=1.4$~mJy, it remains one of the deepest maps ever taken by SCUBA, and it is statistically expected that we would detect one source brighter than 5~mJy (e.g., Coppin et al.\ 2008).
None of the LBGs or the new SMG presented in this region are detected in the radio (Ivison et al.\ 2007) or Xray (Nandra et al.\ 2007), and there is no {\it Spitzer} data covering this area.

%
%
\begin{figure}
\centerline{
\includegraphics[angle=0,width=0.990\hsize]{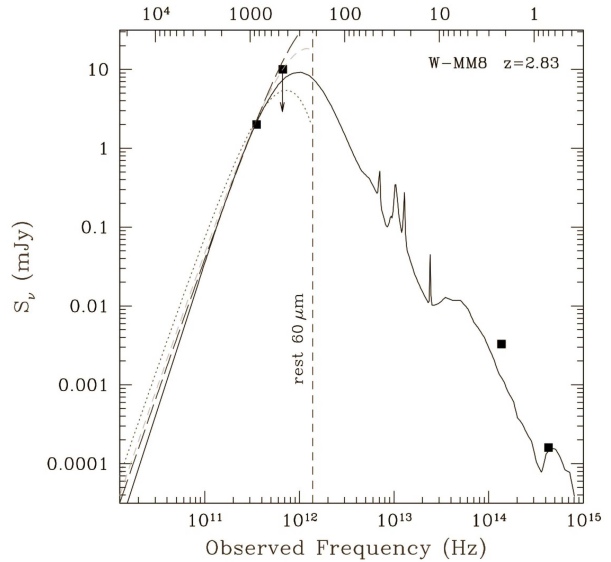}
}
\vskip-1cm
\caption{The spectral  energy distributions of  \mm\ (z=2.83) as
compared  to a representative local starforming galaxy, M82 (solid line).
The top axis shows the observed frame wavelength in microns.
Also plotted in dashed lines are modified blackbody 
spectra for emissivity $\beta=1.5$
and $T_{\rm d}\,{=}\,30$, 50, and 70\,K (dashed lines), 
normalized to the 850\mum\ point.
For nearby galaxies, the IRAS 60\mum\ band has typically been used
to estimate SFRs. We have marked the location of this
band (in the rest-frame) with a vertical dotted line for both objects, as an
indication of the extrapolation used in this study relative to nearby objects.
Optical $R$ and $K$-band photometry comes from Shapley et al.\ (2001).
}
\label{SED}
\end{figure}

%
%
\begin{table*}
\begin{center}
\caption{Predicted and measured (intrinsic) submm flux densities for 
high SFR Lyman break galaxies.
We give the galaxy designation, RA/Dec, redshift,
observed 850\mum\ flux density, 
$R$-band magnitude, 
the SFR derived from the UV-corrected flux,
and the 850\mum\ flux density predicted from the
UV data.}  
\begin{tabular}{@{\extracolsep{-1.5pt}}lllccccc}
\noalign{\medskip}
\hline
{Galaxy} & {RA (J2000)} & {Dec (J2000)} & {$z$} & $\!S_{850}\ ^{\rm a}\!$ & $R_{AB}$ &
$\!\!{\rm SFR}_{\rm UV\!-corr}\ ^{\rm \,b}\!\!$ &
$S_{\rm 850-UV}\ ^{\rm \,c}$ \cr
{} &{} & {} & {} & (mJy) & (mag) &
  $\!\!({\rm M}_\odot\,{\rm yr}^{-1})\!\!$ &
 (mJy)\\
\noalign{\medskip}
\noalign{\bf Unlensed LBGs observed in the SCUBA map} 
West-MM8$^e$  & 14:18:23.90 & +52:23:07.7 & 2.829& $1.98\pm0.48$& 24.13&
 175$\pm$164 & 1.5$\pm$1.1 \\ 
West-MMD11 & 14:18:09.71 & +52:22:08.5 & 2.979& $5.51\pm1.38$& 24.05&
 173$\pm$95 & 1.2$\pm$0.6  \\ 
\noalign{\bf serendipitous SMG discovered in the SCUBA map}
West-SMG1  & 14:18:29.57 & +52:22:54.4 & {--} & $5.83\pm1.39$&  {--}&
 {--} & {--} \\
\noalign{\bf Lensed LBGs from the literature\ $^{\rm d}$}
CosmicEye & 21:35:12.73 & -01:01:42.9 & 3.074& $0.75$& 20.3&
 100 & 2.0 \\ 
MS0451-a& 04:54:10.90 & -03:01:07.0 & 2.911& $0.40$& 22.9&
 10 & 0.2  \\ 
cB58 & 15:14:22.27 & +36:36:25.2 & 2.723& $0.39$ & 20.3&
 24 & 0.6  \\ 
SMM\,J16359& 16:35:54.50 & +66:12:31.0 & 2.517& $0.65$& 23.0&
 9 & 0.3 \\ 
Cosmic Horseshoe& 11:48:33.15 & +19:30:03.5 & 2.379 & $2.03$& 19.0&
80  & 2.3 \\ 
%

\hline
\end{tabular}
\end{center}
\begin{flushleft}
{$^{\rm a}$ $S_{850}$ either measured directly or inferred from CO detections (the latter described in $^d$).}\\
{$^{\rm b}$ Calzetti (1997) attenuation curve corrected, taking into account
Monte Carlo simulations of the photometric errors.  Simple RMS errors
are quoted, while in fact the distributions are skewed (Adelberger \& Steidel 2000).}\\
{$^{\rm c}$ Predicted from the UV colours; see Chapman et al.~2000 and 
Adelberger \& Steidel 2000 for detailed discussion.}\\
{$^{\rm d}$ Properties derived for lensed sources referenced as follows: Cosmic Eye (Coppin et al.\ 2007), MS0451-a (Chapman et al.\ 2002, Borys et al.\ 2004), cB58 (Baker et al.\ 2004), SMM\,J16359 (Kneib et al.\ 2005 -- where we have quoted their central pointing between the multiply imaged galaxy), Cosmic Horseshoe (Belokurov et al.\ 2007, Casey et al.\ in prep).
In two cases where 850$\mu$m measurements were not obtained (Cosmic Eye and Cosmic Horseshoe)
the 850$\mu$m flux was inferred from the CO gas through the relation in Greve et al.\ (2005), where an 
additional uncertainty of $\sim$50\% is introduced by the scatter in SCUBA-bright galaxy properties.}\\
{$^{\rm e}$ The redshift of MM8
appears in Shapley et al. (2001) as 2.829 (as quoted here),  while  $z=2.839$ (in a
table) and $z=2.841$ (in a figure) are given in Steidel et al.\ (2003).}\\
\end{flushleft}
\end{table*}

%
%
\begin{figure}
\centerline{
\includegraphics[angle=0,width=0.80\hsize]{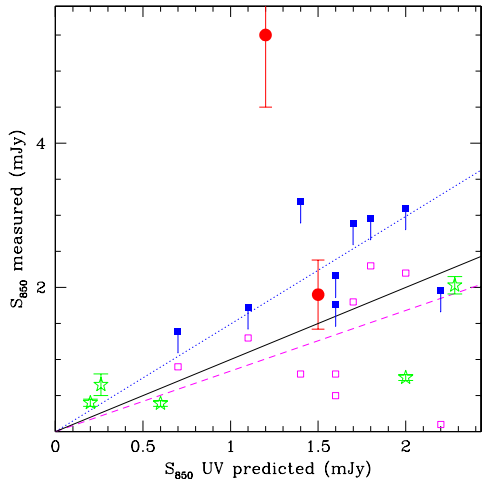}
}
\caption{The measured S$_{\rm 850}$ versus UV-predicted
S$_{\rm 850}$ for $z\sim3$ LBGs, a solid line depicting
the direct $x=y$ relationship. The two detected LBGs (MMD11 and
\mm\ are shown with large red circles. The solid blue squares represent
Bayesian 95\% upper limits to the LBGs presented in Chapman et al.\ (2000),
while the open squares (magenta) show
the measured S$_{850}+1\sigma$ (mJy) points,
where $+1\sigma$ is the minimum
offset to bring all values positive.
The blue (dotted) line represents a fit to the Bayesian 95\% upper 
limits constrained to pass through the point (0,0), 
including the detection of \mm, but excluding the
extreme object Westphal\,MMD11.
The magenta line (dashed) is a fit to the S$_{850}+1\sigma$ points, 
again excluding Westphal\,MMD11. We take this relation as our
high redshift UV/far-IR fiducial for estimation of the LBG
contributions to the submm source count and far-IR background.
Five lensed LBGs with submm measurements are also shown as stars. These LBGs reasonably agree with a one-to-one correspondance of predicted/measured 850$\mu$m flux.
}
\label{SED}
\end{figure}

\section{Results}\label{discussion}

\subsection{\mm\ Luminosity and Dust Content}\label{dustcontent}
Figure~2 presents  the extant  data of \mm\ (optical data points from Shapley et al.\ 2001),  including the
new SCUBA data discussed  in \S~2.  Also plotted is
the  spectral energy  distribution (SED) of a possible local
analog to \mm, the starforming galaxy M82. 
Although still a relatively extreme object, \mm\ represents the faintest
galaxy ever detected with SCUBA, comparable to the faintest object in
the Hubble Deep Field SCUBA observation of Hughes et al.~(1998), HDF850.5 (S$_{850 \mu m}=2.1\pm0.5$~mJy), or the ultra-deep Abell2218 SCUBA map of Knudsen et al.\ (2006),
SMM\,J163602.6+661255 with S$_{850 \mu m}=2.8\pm0.6$~mJy.
This fact emphasizes the difficulty in detecting more typical starforming
galaxies at high redshift in the submm wavelengths with present facilities and achievable sensitivities.

Simply scaling the  observed submm photometry from that of Arp220
implies that \mm\ has  an infrared  luminosity  of ${\rm
L_{IR}\sim2\times10^{12}\  h_{70}^{-2}\ L_\odot}$,  placing  it in the 
ultra-luminous class  of infrared galaxies (Sanders  \& Mirabel 1996).
\mm\  displays  no evidence  of  gravitational lensing,  appearing
point-like in the optical at a PSF-scale $\sim0\scnd7$.
Again scaling from  the Arp220
model,  the resulting  dust  mass  in \mm\  is  ${\rm \sim3\times10^6\
M_\odot}$.   Without more  data over  the SED  of \mm,  this value
possesses a  significant uncertainty, but does  indicate that \mm\
harbours a large quantity of dusty material.
 
In Table~1, we present measured and predicted 850$\mu$m flux
densities along with the UV-derived SFR for the submm-detected LBGs, MM8 and MMD11,
and the lensed LBGs with submm or CO detections are available.
Of the lensed LBGs, it is noteworthy that SMMJ16359 and MS0451-a were discovered on their
submm-properties and not through the LBG selection criterion.
For the two cases where S$_{850}$ was estimated from the CO luminosity, an additional uncertainty of the
order 50\% is likely, as measured directly for the SCUBA-bright population in Greve et al.\ (2005).
%
The UV-predicted S$_{850}$ is highly dependent on the dust parameters
adopted. We have used the empirical dust curve for M82 
which appears to accurately predict S$_{850}$ for \mm, whereas a combination of
different dust emissivity ($\beta$) and dust temperature would
result in significant changes to the predicted submm emission. The uncertainty in this estimate based on the estimated local and high$-z$ dust temperature distributions (e.g., Blain et al.\ 2004) is $\sim$0.3~dex.

\subsection{Implications for the faint submm sources}


A correlation is expected between the rest frame UV continuum and
the far-IR
emission for starforming galaxies, since much of the UV
radiation emitted
by hot O- and B-type stars is absorbed by dust and re-radiated at
far-IR wavelengths in the form of a modified black body.
Meurer et al.~(1999) have shown this to be the case for local
starforming galaxies with bolometric
luminosities L$\sim$10$^{11}$\,L$_\odot$,
with a factor of two scatter in the relation.
However, such a relation can only be expected to hold if the bulk of
the detected UV and far-IR radiation
are emitted from the same location in the galaxy. 
Related is the problem of vastly different optical depth effects
in the two wavelength regimes. 
These issues are verifiable in local galaxies but difficult in high redshift
objects, where typically only the integrated emission is known. 
For extremely luminous local starburst galaxies such as Arp220
(L$\sim$10$^{13}$\,L$_\odot$), the
dominant UV and far-IR components are observed in different spatial
locations, and the integrated UV to far-IR ratio lies significantly
above the Meurer et al.~(1999) correlation (see Goldader et al.\ 2005). 
The UV/far-IR relation
must be taken as unreliable for such hyper-luminous infrared
galaxies.
However, objects like \mm\ lie in the interim luminosity regime
(L$\sim$10$^{12}$\,$\lsun$), for which Reddy et al.\ (2006) have suggested the UV/far-IR relation appears to hold on average based on mid-IR, radio, and X-ray stacking observations.


In Figure~3, we present the measured (or intrinsic) S$_{\rm 850}$ versus UV-predicted
S$_{\rm 850}$ for the LBG sample, depicting
the $x=y$ relationship. The two detected LBGs (Westphal\,MMD11 and
\mm) are highlighted against 
Bayesian 95\% confidence upper limits for literature sources. 
We also show the explicit measurement limits for these LBGs as S$_{850}+1\sigma\times rms$ (mJy).
Two fits are made to unlensed LBGs, one using the Bayesian limits, and one using the S$_{850}+1\sigma$ 
points. The fits are constrained to pass through the origin.
In both cases we exclude MMD11 since its submm luminosity places it in the
class of hyper-luminous objects (Sanders \& Mirabel 1996), and as such
its UV/farIR relation is particularly suspect.
The five lensed LBGs with submm measurements are reasonably described by the fit to the S$_{850}+1\sigma$
data, suggesting the Bayesian limits are likely a conservative estimate of the relation.
%
The relation 
(S$_{850, {\rm measured}}$=0.8$\times$S$_{850, {\rm predicted}}$) is therefore proposed as a 
high redshift UV/far-IR fiducial for estimation of the LBG
contributions to the submm source count.




\section{Discussion}\label{discussion}

Most of the 850$\mu$m background appears to be produced by objects
with S$_{850}\sim$1\,mJy (e.g., Smail et al.\ 2002), and the 
data in this paper are  consistent with the idea that these dusty 1\,mJy sources
which dominate the background have UV properties similar
to LBGs.
With our derived UV/far-IR relation at high redshift,
we calculate a 1-2\,mJy surface density of 3600\,deg$^{-2}$ for LBGs
with expected SFR$>$100$\sfr$ over the redshift range z=1.5-3.5 (using the Reddy et al.\ 2008 luminosity 
function). 
This represents $>$70\% of the measured 1-2\,mJy submm source counts (cf.\ 
Blain et al.~1999, Knudsen et al.\ 2006), and as such suggests that 
optical surveys typically uncover the sources which begin to dominate
the fainter submm source counts.
If all UV-selected $z\sim3$ with SFRs$\sim$100$\sfr$ galaxies are as dusty as the present analysis suggests, then UV-selected populations probably produce
the bulk of the 850\mum\ background (see also Adelberger \& Steidel 2000).  

However, the specific galaxies which dominate the S$_{850}\sim1$~mJy counts is still a
debated issue, where other investigations propose significant contributions
from various classes of galaxies selected at different redshift intervals
(e.g.\ Wehner et al.\ 2002, Webb et al.\ 2003, Knudsen et al.\ 2005, Wang et al.
2006). 
We note firstly that many high redshift galaxies selected at longer wavelengths do have colours
consistent with the LBG selection (e.g., Reddy et al.\ 2008), including a substantial fraction of the hyper-luminous
SMGs (Chapman et al.\ 2005), and thus one should expect an overlap between the different
galaxy classes that are estimated to contribute signicantly. 
The second obvious problem with these analyses is the large ($\sim15$\arcsec) confusion limited beamsizes
of submm/mm measurements, allowing the galaxies of different types which may well cluster together to
be collectively measured in the same 850$\mu$m beam.
Of note is the low average 850$\mu$m flux for LBGs inferred by Chapman et
al.\ (2000), Peacock et al.\ (2000), and Webb et al.\ (2003), $<1$~mJy. Individual objects in these samples likely have SFRs substantially lower than the UV-inferred SFR due to the photometric uncertainties described in the introduction.
Finally, we note the indirect estimates of the SFR. Without direct 
constraint of the far-IR peak for high redshift, luminous galaxies, there remains substantial uncertainty in 
estimating how any particular population contributes to the far-IR backgrounds.
 




The discovery, confirmation, and full characterization of $z\sim2-2.5$ evolved galaxies with no ongoing star formation (e.g., Kriek et al.\ 2008, van Dokkum et al.\ 2008) demands a sizable progenitor population to rapidly form the stars at $z>3$. The submm-luminous galaxy population (S$_{850 \mu m}>5$mJy) for the most part cannot fulfill this role as they lie predominantly {\it at} $z\sim2$ (e.g., Chapman et al.\ 2005,
Ivison et al.\ 2007, Clements et al.\ 2008), coeval with this old evolved population.
These submm-selected galaxies have a rapidly declining luminosity function beyond $z>2.5$, dropping in volume density over $2\times$ from $z=2$--3, and likely a factor $\sim$10 from $z=2$--4
(Chapman et al.\ 2005, Pope et al.\ 2006). However galaxies with S$_{850 \mu m}\sim$1--2\,mJy could remain be numerous at $z>3$ (e.g., Adelberger \& Steidel 2000) and would still represent an ultra-luminous population.

While we do not extrapolate submm properties explicitly from this single submm detection of a {\it typical} LBG at $z\sim3$, 
we state for reference the expected volume density of $z\sim3$ LBGs with likely S$_{850\mu m}>$1~mJy (total infrared luminosities
L$_{\rm IR}>10^{12}$~L$_\odot$) from Reddy et al.\ (2008): 
$\Phi \sim 3\times10^{-4} {\rm h^3_{0.7} Mpc^{-3}}$, which is very similar to their estimate for $z\sim2$ indicating little redshift evolution. Contrast bright submm-selected galaxies which would only have $\Phi < 10^{-6}{\rm h^3_{0.7} Mpc^{-3}}$ at $z$\gta3 (Chapman et al.\ 2005).


Measuring the submm fluxes of large samples of known $z=2-4$ galaxies will be a cornerstone goal of SCUBA2 (Holland et al.\ 2003), where vast confusion-limited surveys at both 850$\mu$m and 450$\mu$m will be achieved.
This will allow the extension of this study to large samples of galaxies, although the issue of confusion noise
will persist at 850$\mu$m.
As shown in Fig.~2, the limit we achieved for MM8 at 450$\mu$m for \mm\ is not quite deep enough to discriminate
between different dust temperatures (or SED shapes in general). However, the confusion-limited 
450$\mu$m sensitivities ($\sim$0.5~mJy rms) listed in the goals of the deep  SCUBA2 Legacy programmes
\footnote{http://www.jach.hawaii.edu/JCMT/surveys/Cosmology.html} are more than sufficient to completely characterize dust temperature distributions even for $z>3$ galaxies, and the reasons for 
the submm detectability raised in \S~1 will become clear.
E-MERLIN\footnote{http://www.merlin.ac.uk/e-merlin} and E-VLA\footnote{http://www.aoc.nrao.edu/evla} will also be able to detect these populations directly at 1.4\,GHz, and in addition
constrain their radio sizes, potentially well in advance of being able to study their sizes directly in the submm with
ALMA\footnote{http://almaobservatory.org}.
%
%
Finally, followup of unlensed LBGs (using the IRAM, Plateau de Bure Interferometer) in molecular CO gas has met with four failures and no successes
(Baker et al.\ 2004b, Tacconi et al.\ 2008). MM8 represents an obvious candidate for CO followup, 
however based on the current lack of success, it is likely that ALMA will be required for full characterization of these galaxies dominating the submm background.







\section{Conclusions}\label{conclusions}
We  have  presented submm observations  of the $z=2.83$ LBG
\mm, nominally the deepest SCUBA observation ever taken of a distant galaxy.
These observations  confirm that  this  LBG has a large star formation rate
($\sim200$\,M$_\odot$/yr) as indicated by the dust-corrected UV.
\mm\ is associated with a large 
quantity  of  dust, with  an  inferred  infrared  luminosity of  ${\rm
\sim2\times10^{12}~h_{70}^{-2}~L_\odot}$.
%

Despite the large dust mass, enough
of the UV remains in view to provide a clear indication of the
expected submm flux. This contrasts with typical submm-luminous galaxies where high resolution radio and mm-interferometry (Chapman et al.\ 2004b, Tacconi et al.\ 2006, 2008) have demonstrated that the 
bolometrically luminous region can lie significantly offset (several kpc) from UV-luminous regions. 
This suggests that  for  less luminous $z\sim3$ galaxies like
\mm,  the dust distribution may not hinder the
UV/far-IR relation to the same extent. 
The observations presented here also suggest that
individual LBGs from the sample of Chapman et al.~(2000) 
could be similar in submm properties to \mm\ ($\sim$1--2\,mJy), and SCUBA2 surveys may regularly detect these galaxies in its confusion limited wide field surveys.

\section*{Acknowledgements}
We thank an anonymous referee for helpful comments which improved the manuscript.
SCC acknowledges support from the Canadian Space Agency and NSERC.
CMC thanks the Gates-Cambridge Trust for support.
The James
Clerk Maxwell Telescope  is operated by The Joint  Astronomy Centre on
behalf of the  Particle Physics and Astronomy Research  Council of the
United Kingdom, the  Netherlands Organization for Scientific Research,
and the National Research Council of Canada.
We acknowledge the use of the Canadian Astronomy Data Centre, 
which is operated by the Herzberg Institute of Astrophysics, National
Research Council of Canada.

\label{lastpage}

\end{document}